\documentclass[prd,superscriptaddress,a4paper,showpacs,showkeys,10pt,nofootinbib]{revtex4}
\usepackage{graphicx}
\usepackage{graphics}
\usepackage{epsfig}
\usepackage{dcolumn}
\usepackage{bm}
\usepackage{multirow}
\usepackage{tabularx}
\usepackage{hyperref}
\usepackage{commath}
\usepackage{epstopdf}
\usepackage[T1]{fontenc}
\usepackage{geometry}
\usepackage{color}

\def\lsim{\raise0.3ex\hbox{$<$\kern-0.75em\raise-1.1ex\hbox{$\sim$}}}

\def\gsim{\raise0.3ex\hbox{$>$\kern-0.75em\raise-1.1ex\hbox{$\sim$}}}

\def\pom{{I\!\!P}}

\newcommand{\be}{\begin{equation}}

\newcommand{\ee}{\end{equation}}

\def\beq{\begin{equation}}

\def\eeq{\end{equation}}

\def\beqa{\begin{eqnarray}}

\def\eeqa{\end{eqnarray}}

\newcommand{\rd}{\mbox{\boldmath $\Delta$}}

\newcommand{\ba}{\begin{eqnarray}}

\newcommand{\ea}{\end{eqnarray}}

\newcommand{\rr}{\mbox{\boldmath $r$}}

\newcommand{\rb}{\mbox{\boldmath $b$}}

\def\gappeq{\mathrel{\rlap {\raise.5ex\hbox{$>$}}

{\lower.5ex\hbox{$\sim$}}}}

\def\lappeq{\mathrel{\rlap{\raise.5ex\hbox{$<$}}

{\lower.5ex\hbox{$\sim$}}}}

\def\Toprel#1\over#2{\mathrel{\mathop{#2}\limits^{#1}}}

\def\pom{{I\!\!P}}

\begin{document}

\title{Exclusive vector meson production in electron -- ion collisions \\ at the EIC, LHeC and FCC--$eh$}
\author{Victor P. Gon\c{c}alves}
\email[]{barros@ufpel.edu.br}
\affiliation{High and Medium Energy Group, Instituto de F\'{\i}sica e Matem\'atica,  Universidade Federal de Pelotas (UFPel)\\
Caixa Postal 354,  96010-900, Pelotas, RS, Brazil. \\
}

\author{Daniel E. Martins}
\email[]{dan.ernani@gmail.com}
\affiliation{Instituto de F\'isica, Universidade Federal do Rio de Janeiro (UFRJ), 
Caixa Postal 68528, CEP 21941-972, Rio de Janeiro, RJ, Brazil}

\author{Celso R. Sena}
\email[]{celsorodriguessena@gmail.com}
\affiliation{High and Medium Energy Group, Instituto de F\'{\i}sica e Matem\'atica,  Universidade Federal de Pelotas (UFPel)\\
Caixa Postal 354,  96010-900, Pelotas, RS, Brazil. \\
}

\begin{abstract}
The exclusive vector meson production in electron -- ion collisions for the energies of the future colliders is investigated. We present predictions for the coherent and incoherent $\phi$ and $J/\psi$ production in $eAu$ collisions considering the possible states of nucleon configurations in the nuclear wave function and taking into account of the non - linear corrections to the QCD dynamics.
The cross sections and  transverse momentum distributions are estimated assuming the energies of the Electron - Ion Collider (EIC), Large Hadron Electron Collider (LHeC) and Future Circular Collider (FCC -- $eh$). Our results indicate that a future experimental analysis of these processes can shed light on the  modeling of the gluon saturation effects and constrain the description of the QCD dynamics at high energies.
\end{abstract}



\maketitle

\vspace{1cm}

Electron - hadron colliders are the ideal facilities to improve our understanding of the strong interactions theory -- the Quantum Chromodynamics (QCD) -- in the high energy regime, where the gluons play a dominant role in the hadron structure and non-linear (saturation) effects are expected to become important \cite{hdqcd}. The search for these effects is one of the major motivations for the construction of the Electron - Ion Collider (EIC) in the USA \cite{eic}, recently approved, as well as for the proposal of future electron -- hadron colliders at CERN \cite{lhec}. These colliders are expected to allow the investigation of the hadronic structure with unprecedented precision to inclusive and diffractive observables. 
In particular, electron -- nucleus collisions are considered ideal to probe the saturation regime \cite{eic_general}.  The higher parton densities in the nuclear case  enhance  by a factor $ \propto A^{\frac{1}{3}}$ the nuclear saturation scale, $Q^2_{s,A}$, which determines the onset of non-linear effects in the QCD dynamics. Over the last years, such expectation have motivated an intense phenomenology about the implications of the gluon saturation effects in the observables \cite{eic,lhec}. Such studies have demonstrated that the analysis of diffractive events can be considered a smoking gun of the gluon saturation effects in $eA$ collisions 
\cite{erike_ea2,vmprc,Caldwell,Lappi_inc,Toll,armestoamir,diego,Lappi:2014foa,Mantysaari:2016ykx,Mantysaari:2016jaz,Diego1,contreras,Luszczak:2017dwf,Mantysaari:2017slo,Diego2,Bendova:2018bbb,cepila,Lomnitz:2018juf,Mantysaari:2019jhh}. Such events are predicted to contribute with half of the total cross section in the asymptotic limit of very high energies, with the other half being formed by all inelastic processes \cite{Nikolaev,simone2,Nik_schafer,Kowalski_prc}, and the associated observables depend on the square of the scattering amplitude, which makes them  strongly sensitive to the underlying QCD dynamics.

\begin{figure}[t]
\begin{tabular}{cc}
 {\includegraphics[width=0.5\textwidth]{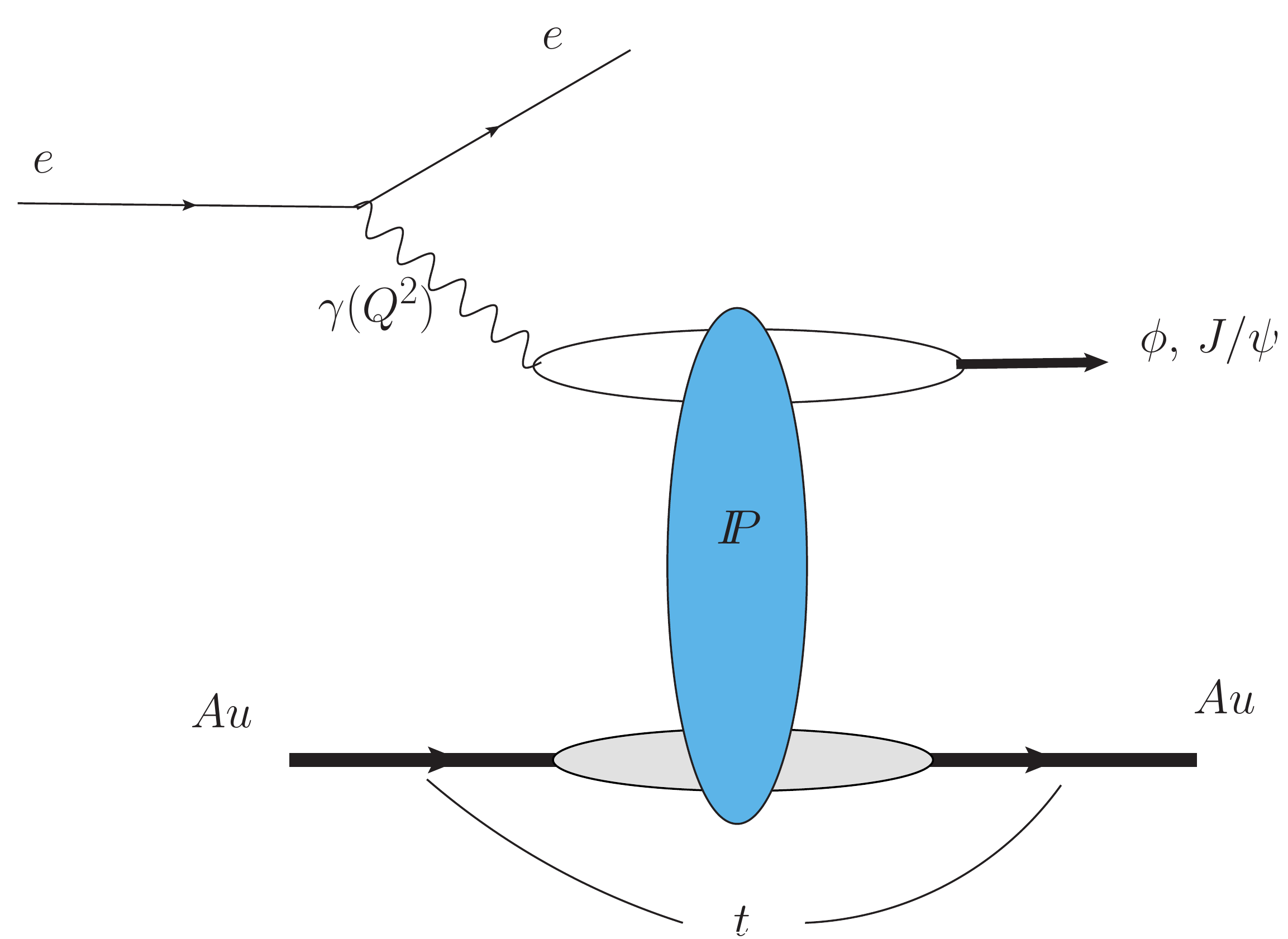}} & 
{\includegraphics[width=0.5\textwidth]{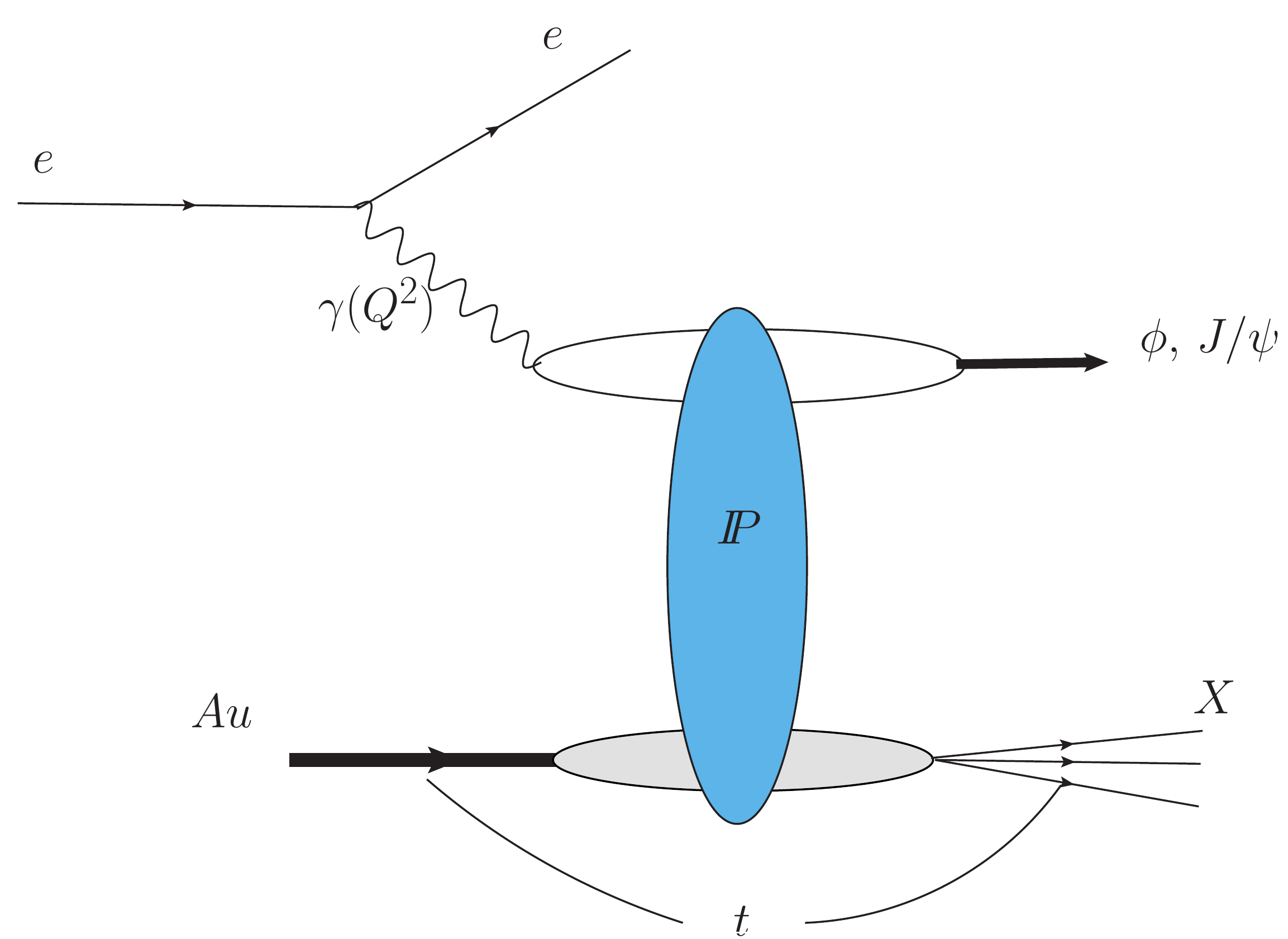}} \\ 
 (a) & (b)  
\end{tabular}                                                                                                                       
\caption{Typical diagrams for the (a) coherent and (b) incoherent vector meson production in electron -- ion collisions.}
\label{fig:diagrama}
\end{figure}

One of the most promising observables to probe the gluonic structure of nuclei and the high - density regime of QCD is the 
exclusive vector meson production off large nuclei 
in coherent and incoherent interactions, represented in Figs. \ref{fig:diagrama} (a) and (b), respectively. Both processes are characterized by the presence of a rapidity gap in the final state, due to the color singlet exchange. However, 
if the nucleus scatters elastically,  the process is called coherent production. On the other hand,  if the nucleus scatters inelastically, the process is denoted incoherent production and  one sums over all final states of the target nucleus,
except those that contain particle production. 
In the color dipole formalism, the scattering amplitude  can be factorized in terms of the fluctuation of the  virtual photon into a $q \bar{q}$ color dipole, the dipole-nucleus scattering by a color singlet exchange ($\pom$)  and the recombination into the exclusive final state, being given by (See, e.g. Ref. \cite{kmw})
\begin{eqnarray}
 {\cal A}_{T,L}({x},Q^2,\Delta)  =  i
\,\int d^2\rr \int \frac{dz}{4\pi} \int \, d^2\rb \, e^{-i[\rb -(1-z)\rr].\rd}
 \,\, (\Psi^{V*}\Psi)_{T,L}  \,\,\frac{d\sigma_{dA}}{d^2\rb}({x},\rr,\rb)
\label{amp}
\end{eqnarray}
where $T$ and $L$ denotes the transverse and longitudinal polarizations of the virtual photon, $\Delta = \sqrt{-t}$ is the momentum transfer, $Q^2$ is the photon virtuality and $x = (M^2 + Q^2 -t)/(W^2+Q^2)$, with  $W$ being the center of mass energy of the virtual photon -- nucleus system and $M$  the mass of the vector meson. 
The variables  $\rr$ and $z$ are the dipole transverse radius and the momentum fraction of the photon carried by a quark (an antiquark carries then $1-z$), respectively, and   $\rb$ is the impact parameter of the dipole relative to the target. Moreover,    $(\Psi^{V*}\Psi)_i$ denotes the wave function overlap between the virtual photon and the vector meson wave functions and ${d\sigma_{dA}}/{d^2\rb}$  is the dipole-nucleus cross section (for a dipole at  impact parameter $\rb$) which encodes all the information about the hadronic scattering, and thus about the non-linear and quantum effects in the hadron wave function \cite{hdqcd}. The differential cross sections for the coherent and incoherent interactions are given by
\begin{equation}\label{eq:xsec-coh}
  \left.\frac{d\sigma^{\gamma^* A \rightarrow V \,A}}{dt}\right|^{coh}_{T,L} =
  \frac{1}{16\pi}\left| \left\langle \mathcal{A}_{T,L}(x,Q^2, \Delta) \right\rangle \right|^2\,\,,
\end{equation}
and 
\begin{equation}\label{eq:xsec-inc}
  \left.\frac{d\sigma^{\gamma^* A \rightarrow V\,X}}{dt}\right|^{inc}_{T,L} = \frac{1}{16\pi}
  \left(  \left\langle\left|  \mathcal{A}_{T,L}(x,Q^2, \Delta)  \right|^2 \right\rangle  - \left| \left\langle \mathcal{A}_{T,L}(x, Q^2, \Delta) \right\rangle \right|^2\right),
\end{equation}
where $\left\langle ... \right\rangle$  represents the average over the configurations of the nuclear wave function and   $X = A^*$ is the dissociative state generated in the incoherent interaction. As demonstrated in Refs. \cite{Toll,Mantysaari:2016ykx,Mantysaari:2016jaz,cepila}, the coherent and incoherent  productions probe different aspects of the  gluon distribution of the target. One has that  in the coherent case, the average is performed at the
 level of the scattering amplitude. It implies that the averaged density profile of the gluon density is probed in these processes.  
  On the other hand, in the incoherent case,   
  the average over configurations is at the cross section level, which implies that the resulting incoherent cross section is proportional to the variance of the amplitude with respect to the nucleon configurations of the nucleus. As a consequence, the incoherent processes measure the fluctuations of the gluon density inside the nucleus. During the last years, several studies of the coherent and incoherent vector meson production in electron -- ion and ultraperipheral heavy ion collisions were performed considering different final states and distinct treatments for the dipole -- nucleus cross section and average over nucleon configurations (See, e.g. Refs. 
\cite{Toll,contreras,cepila,Mantysaari:2019jhh,nos_periferica}). Our goal in this letter is to extend these previous studies, presenting numerical results computed using two distinct models for the QCD dynamics for the $J/\psi$ and $\phi$ production in the kinematics relevant for the EIC, LHeC and FCC-$eh$. We will present predictions for the total cross sections and  transverse momentum distributions, obtained taking into account of the saturation effects, and the results will be compared with those derived disregarding these effects. Predictions for the electroproduction of vector mesons in electron - ion collisions at the LHeC and FCC-$eh$ energies are presented for the first time in the literature.

Initially, let's present the main ingredients in our calculations of the coherent and incoherent cross sections. As in previous calculations,   
  the wave function overlap functions, $(\Psi^{V*}\Psi)_i$,  will be described using the Boosted Gaussian model (For details see e.g. Ref. \cite{run2}).
  Moreover,  we will describe the dipole - nucleus cross section using the
Glauber-Gribov formalism \cite{glauber,gribov,mueller}, which implies that ${d\sigma_{dA}}/{d^2\rb}$ is given by
\begin{eqnarray}
\frac{d\sigma_{dA}}{d^2\rb} = 2\,\left( 1 - \exp \left[-\frac{1}{2}  \, \sigma_{dp}(x,\rr^2) \,T_A(\rb)\right]\right) \,\,,
\label{enenuc}
\end{eqnarray}
where $\sigma_{dp}$ is the dipole-proton cross section and $T_A(\rb)$ is  the nuclear profile function. 
  Moreover, as in Ref. \cite{Toll}, the dipole - proton cross section  will be given by
\begin{equation}
\sigma_{dp}(x,\rr^2) = \frac{\pi^2 r^{2}}{ N_{c}} \alpha_{s}(\mu^{2}) \,\,xg\left(x, \mu^2 = \frac{4}{r^{2}} + 
\mu_{0}^{2}\right) \,\,\,
\end{equation}
where the gluon distribution evolves via DGLAP equation, with the initial condition at $\mu_{0}^{2}$   taken to be $
xg(x,\mu_{0}^{2}) =  A_{g}x^{-\lambda_{g}} (1-x)^{5.6}$.  
In this work, we 
assume the parameters $ A_g, \lambda_g$ and $\mu_0^2$ obtained in Ref. \cite{kmw} for the IP-SAT model.  
We will denote by b - Sat the predictions derived using  Eq. (\ref{enenuc}) as input in the calculations. For comparison, we also will present predictions derived disregarding the saturation effects, with 
the dipole - nucleus cross section being given by:
\begin{eqnarray}
\frac{d\sigma_{dA}}{d^2\rb} =  \sigma_{dp}(x,\rr^2) \,T_A(\rb) \,\,.
\label{enenuc_lin}
\end{eqnarray}
 The associated predictions will be denoted by b - Non Sat hereafter.  In our calculations we also will include the skewedness correction by multiplicating the coherent and incoherent cross sections by the factor $R_g^2$ as given in Ref. \cite{Shuvaev:1999ce}.
Moreover,  the nuclear profile $T_A(\rb)$ will be described taking into account of all possible states of nucleon configurations in the nuclear wave function.
 We will assume  that each nucleon in the nucleus  has a Gaussian profile of width $B_G$ centered at random positions sampled from a Woods-Saxon nuclear profile as follows \cite{Toll,cepila}
\begin{equation}\label{eq:Ths0}
  T_A(\rb) = \frac{1}{2\pi B_G} \sum_{i=1}^{A} \exp\left[ - \frac{(\rb - \rb_i)^2}{2B_G} \right] \,\,,
\end{equation}
with $B_G = 4.0$ GeV$^2$, as determined in Ref. \cite{kmw} through fits to HERA data.
 The numerical calculations will be performed using the Sar{\it t}re event generator proposed in Ref. \cite{Toll} and detailed in Ref. \cite{sartre}.  
  In order to perform the averages present in the coherent and incoherent cross sections, we have considered 500 distinct nucleon configurations. As demonstrated in \cite{Toll}, this number of configurations is enough to obtain a good description of the cross sections for $|t| \le 0.08$ GeV$^2$, which is the range of interest in our study.

\begin{table}[t]
\centering
\begin{tabular}{||c|c|c||}
\hline 
\hline
{\bf Collider}  &  Electron Energy  & Hevy Ion Energy    \\ \hline
EIC  &  21 GeV & 100 GeV/nucleon      \\ \hline 
LHeC  &  60 GeV & 2800 GeV/nucleon      \\ \hline 
FCC - $eh$  &  60 GeV & 7885 GeV/nucleon      \\ 
\hline 
\hline  
\end{tabular} 
\caption{Electron and heavy ion energies considered in our analysis of electron -- ion collisions at the EIC, LHeC and FCC - $eh$.}
\label{tab:ene}
\end{table}

  In what follows we will analyze the  photoproduction ($Q^2 \approx 0$) and electroproduction ($Q^2 \ge 1$ GeV$^2$) of vector mesons in electron - ion collisions,   presenting predictions for the total cross sections and transverse momentum distributions. We consider three different configurations for the electron and heavy ion energies, summarized in Table \ref{tab:ene}, which correspond to those planned for the EIC, LHeC and FCC - $eh$. As we are interested in the energy dependence of our predictions, we will assume for all colliders that $A = 197$. The atomic  mass number dependence has been investigated e.g. in Ref. \cite{Mantysaari:2017slo}, which we refer the reader for a more detailed discussion.  Moreover, we have selected the	 events in which $|t| \le 0.08$ GeV$^2$. 
Our focus is to estimate the impact of the saturation effects  on the vector meson production. Such effects are predicted to be dominant in the kinematical range where $Q_{s,A} \gtrsim \mu$, with $\mu$ being the hard scale of the process.  In the case of the vector meson production, one has that  $\mu \propto \sqrt{M^2 + Q^2}$. Moreover, for a fixed $Q^2$,   the main contribution for the overlap function $(\Psi^{V*}\Psi)_i$ comes from dipoles with  larger values of $\rr$ for the $\phi$  meson than for $J/\psi$ \cite{run2}. Therefore, the light and heavy meson states probe the dipole - nucleus cross section at different values of $\rr$. So, studying these two  
final states we are mapping different configurations of the dipole size, which 
probe different regimes of the QCD dynamics. In particular, we expect that the production of $\phi$ mesons is more sensitive to saturation effects than heaviers mesons as $J/\psi$, especially for smaller values of $Q^2$. In addition, we also expect that the impact of these effects decreases with the increasing of the photon virtuality.  
  In Tables \ref{tab:cross_psi} and \ref{tab:cross_phi} we present our results for the coherent and incoherent $J/\psi$ and $\phi$ production, respectively, considering distinct ranges of  $Q^2$.  The cross sections increase with the energy and decrease at larger virtualities. As expected, the impact of the saturation effects is larger for the $\phi$ production than for $J/\psi$, mainly in the range $Q^2 \le 1$ GeV$^2$, with the gluon saturation effects decreasing the magnitude of the cross sections. In particular, for the exclusive $\phi$ production, the b-Sat and b-Nonsat predictions differ by a factor $\ge 10$ for  $Q^2 \le 1$ GeV$^2$. On the other hand, for the $J/\Psi$ production,  the predictions are similar. These results are expected, since the gluon saturation effects are predicted to suppress the contribution of the large size dipoles, which are dominant in the $\phi$ case in the photoproduction regime, but contribute less for the  $J/\Psi$ production.    
  One also has that the coherent process is dominant,  which is expected, since the coherent production is characterized by a sharp
forward diffraction peak, being much larger than the incoherent one for small values of $|t|$ (see below).
  We predict cross sections of  the order of nb, which implies a large number of events per year at the EIC / LHeC / FCC - $eh$, given the high luminosity expected for these accelerators \cite{eic,lhec}. As a consequence, we expect that a future  analysis of the coherent and incoherent processes will be, in principle, feasible. Our results indicate that such study can be useful to discriminate between the b-Sat and b-Non Sat scenarios.

\begin{table}[t]
\centering
\begin{tabular}{|c||c|c||c|c||c|c|}\hline 
{\bf $J/\psi$} &   \multicolumn{2}{|c||}{\bf EIC} &    \multicolumn{2}{|c||}{\bf LHeC} &     \multicolumn{2}{|c|}{\bf FCC -- eh}      \\ \hline         
{\bf Dipole Model}  &  b-Sat & b-Non Sat  &  b-Sat & b-Non Sat  &  b-Sat & b-NonSat  \\ \hline        
\hline
{\bf Coherent} &  & & & & & \\\hline
{$Q^{2} \le 1.0$ GeV$^{2}$}  &  521.8 & 716.9     &  1095.5  & 1563.0       &  1103.0  & 1578.5      \\ \hline 
{$1.0 \le Q^{2} \le 10.0$ GeV$^{2}$}  &  61.2 & 81.2     &  138.5  & 188.0       &  139.0  & 188.7      \\ \hline 
{$5.0 \le Q^{2} \le 10.0$ GeV$^{2}$}  &  8.8 & 11.1     &  21.7  & 28.0       &  22.0  & 28.4      \\ \hline 
\hline
{\bf Incoherent}&  & & & & & \\\hline
{$Q^{2} \le 1.0$ GeV$^{2}$}  &  39.3 & 82.5     &  77.3  & 150.3       &  78.1  &  151.8      \\ \hline 
{$1.0 \le Q^{2} \le 10.0$ GeV$^{2}$}  &  4.8 & 8.8     &  10.1  & 17.5       &  10.2  &  17.7      \\ \hline 
{$5.0 \le Q^{2} \le 10.0$ GeV$^{2}$}  &  0.7 & 1.1     &  1.6  & 2.5       &  1.7  &  2.6      \\ \hline 
\hline  
\end{tabular} 
\caption{Cross sections, in nb, for the coherent and incoherent $J/\psi$ production in $eAu$ collisions at EIC, LHeC and FCC - $eh$. Predictions for  $|t| \le 0.08$ GeV$^2$ and distint ranges of the photon virtuality $Q^2$.}
\label{tab:cross_psi}
\end{table}

\begin{table}[t]
\centering
\begin{tabular}{|c||c|c||c|c||c|c|}\hline 
{\bf $\phi$} &   \multicolumn{2}{|c||}{\bf EIC} &    \multicolumn{2}{|c||}{\bf LHeC} &     \multicolumn{2}{|c|}{\bf FCC -- eh}      \\ \hline         
{\bf Dipole Model}  &  b-Sat & b-Non Sat  &  b-Sat & b-Non Sat  &  b-Sat & b-NonSat  \\ \hline        
\hline
{\bf Coherent} &  & & & & & \\\hline
{$Q^{2} \le 1.0$ GeV$^{2}$}  &  24544.4 & 304383.0     &  39022.0  & 453970.0       &  39870.0  & 462055.0      \\ \hline 
{$1.0 \le Q^{2} \le 10.0$ GeV$^{2}$}  &  407.1 & 2308.4     &  726.7  & 3832.0       &  732.3  & 3874.1      \\ \hline 
{$5.0 \le Q^{2} \le 10.0$ GeV$^{2}$}  &  16.3 & 48.5     &  33.0  & 92.4       &  33.3  & 91.7      \\ \hline 
\hline
{\bf Incoherent} &  & & & & & \\\hline
{$ Q^{2} \le 1.0$ GeV$^{2}$}  &  4938.3 & 73192.1     &  5564.0  & 83060.0       &  5667.2  &  84349.1      \\ \hline 
{$1.0 \le Q^{2} \le 10.0$ GeV$^{2}$}  &  71.4 & 528.6     &  87.0  & 635.6       &  87.7  &  645.8      \\ \hline 
{$5.0 \le Q^{2} \le 10.0$ GeV$^{2}$}  &  2.1 & 8.3     &  3.1  & 11.6       &  3.2  &  12.0      \\ \hline 
\hline  
\end{tabular} 
\caption{Cross sections, in nb, for the coherent and incoherent $\phi$ production in $eAu$ collisions at EIC, LHeC and FCC - $eh$. Predictions for  $|t| \le 0.08$ GeV$^2$ and distint ranges of the photon virtuality $Q^2$.}
\label{tab:cross_phi}
\end{table}

 \begin{figure}[t]
\begin{tabular}{ccc}
 {\includegraphics[width=0.35\textwidth]{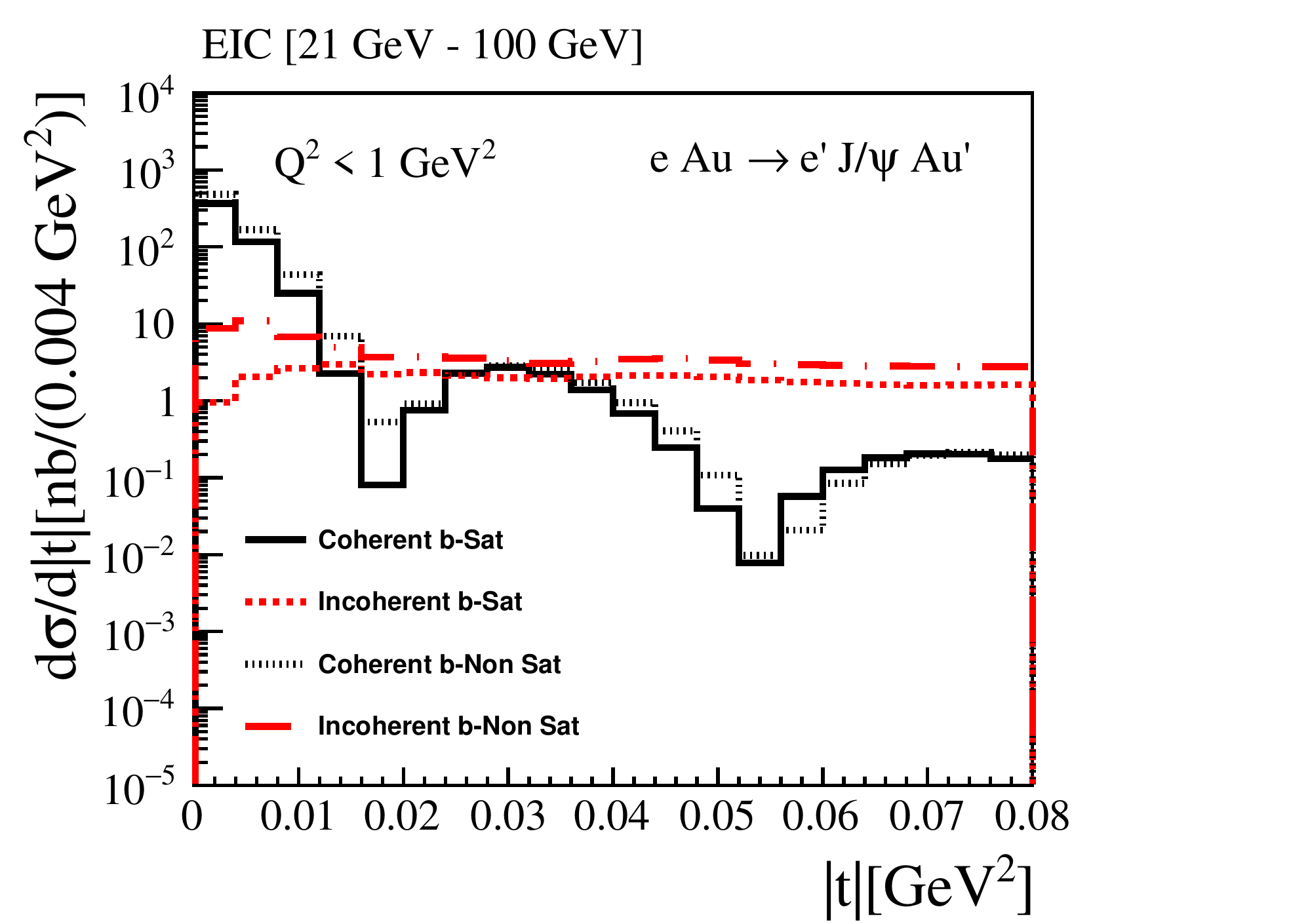}} & 
{\includegraphics[width=0.35\textwidth]{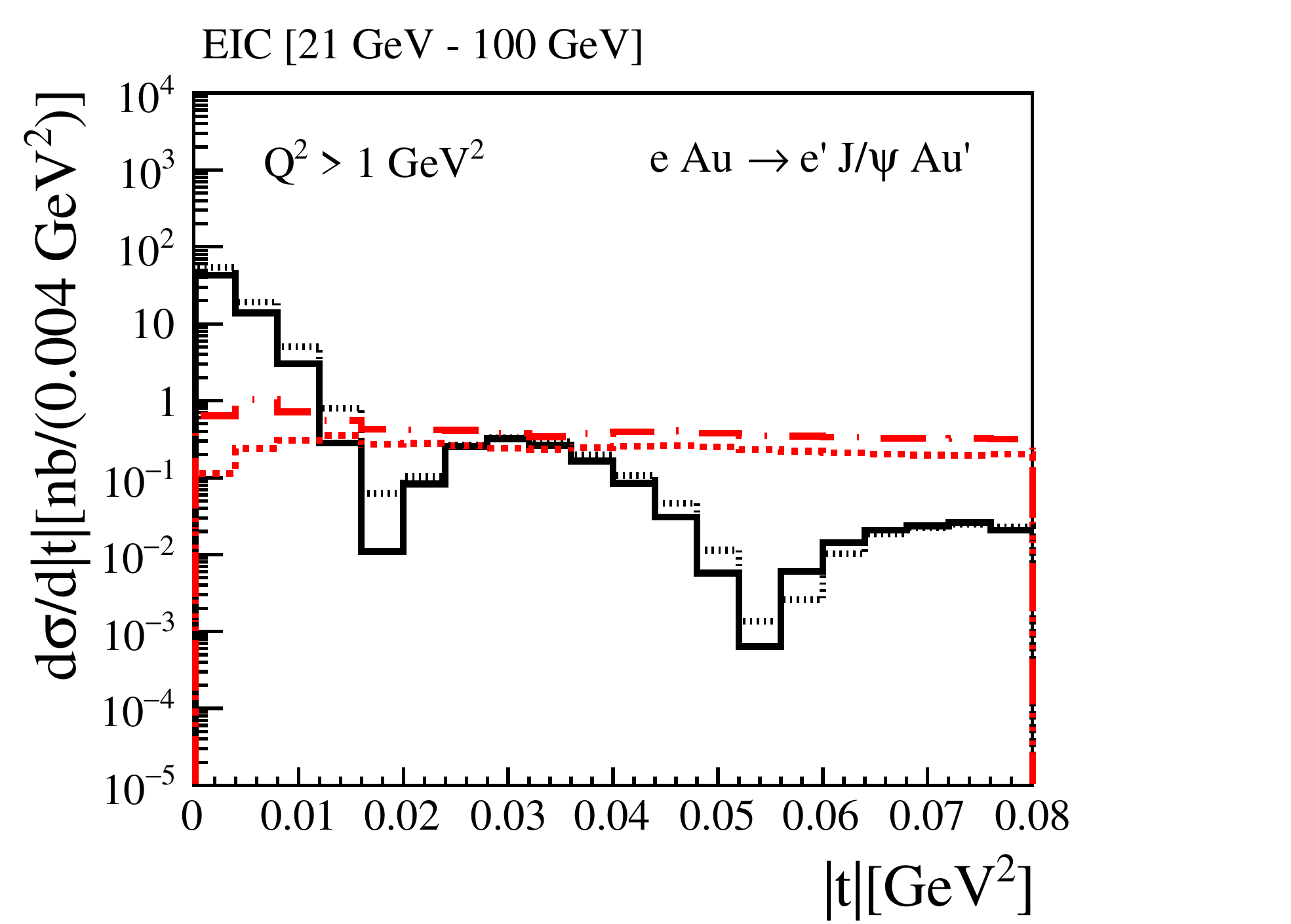}} & {\includegraphics[width=0.35\textwidth]{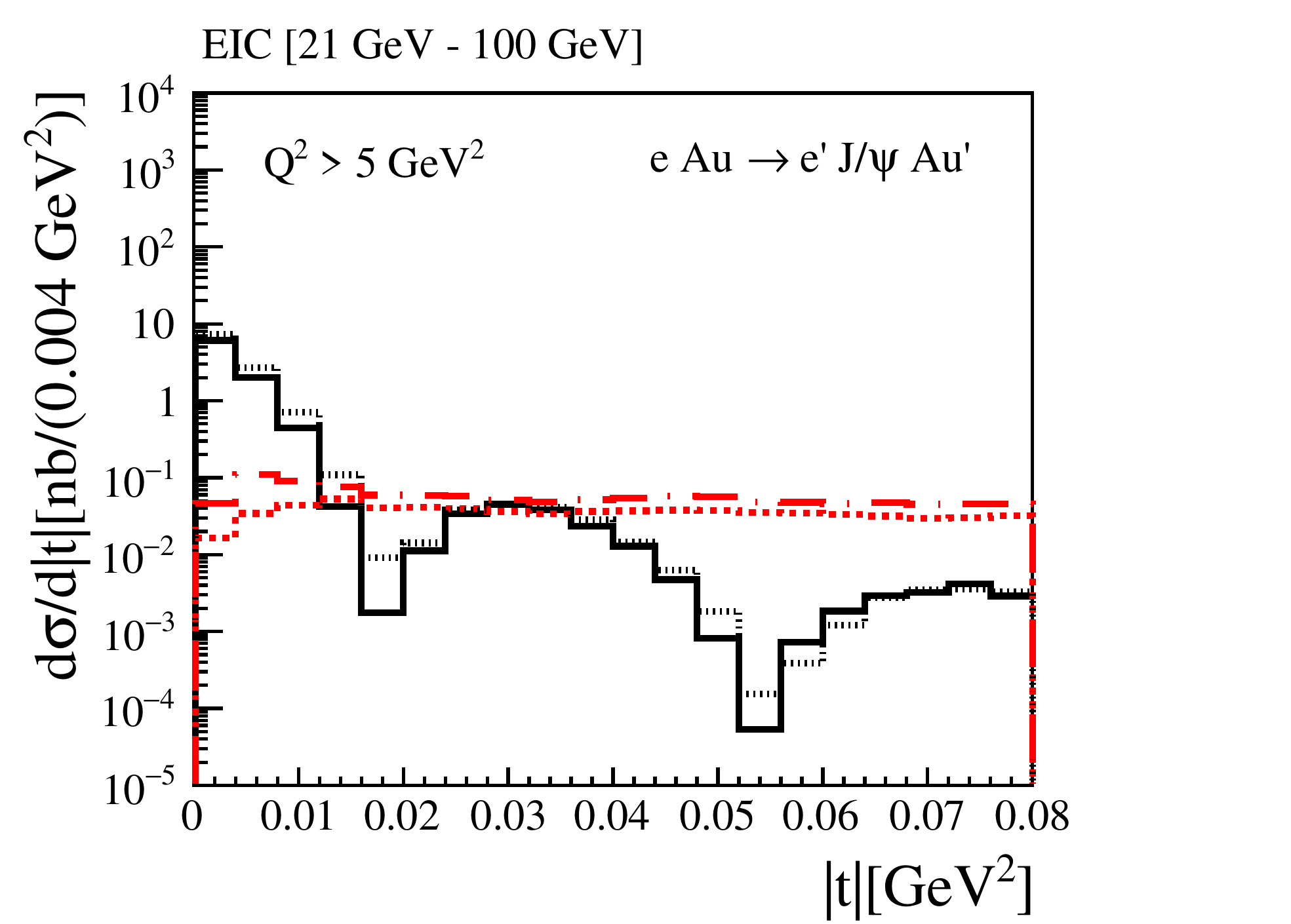}} \\
 {\includegraphics[width=0.35\textwidth]{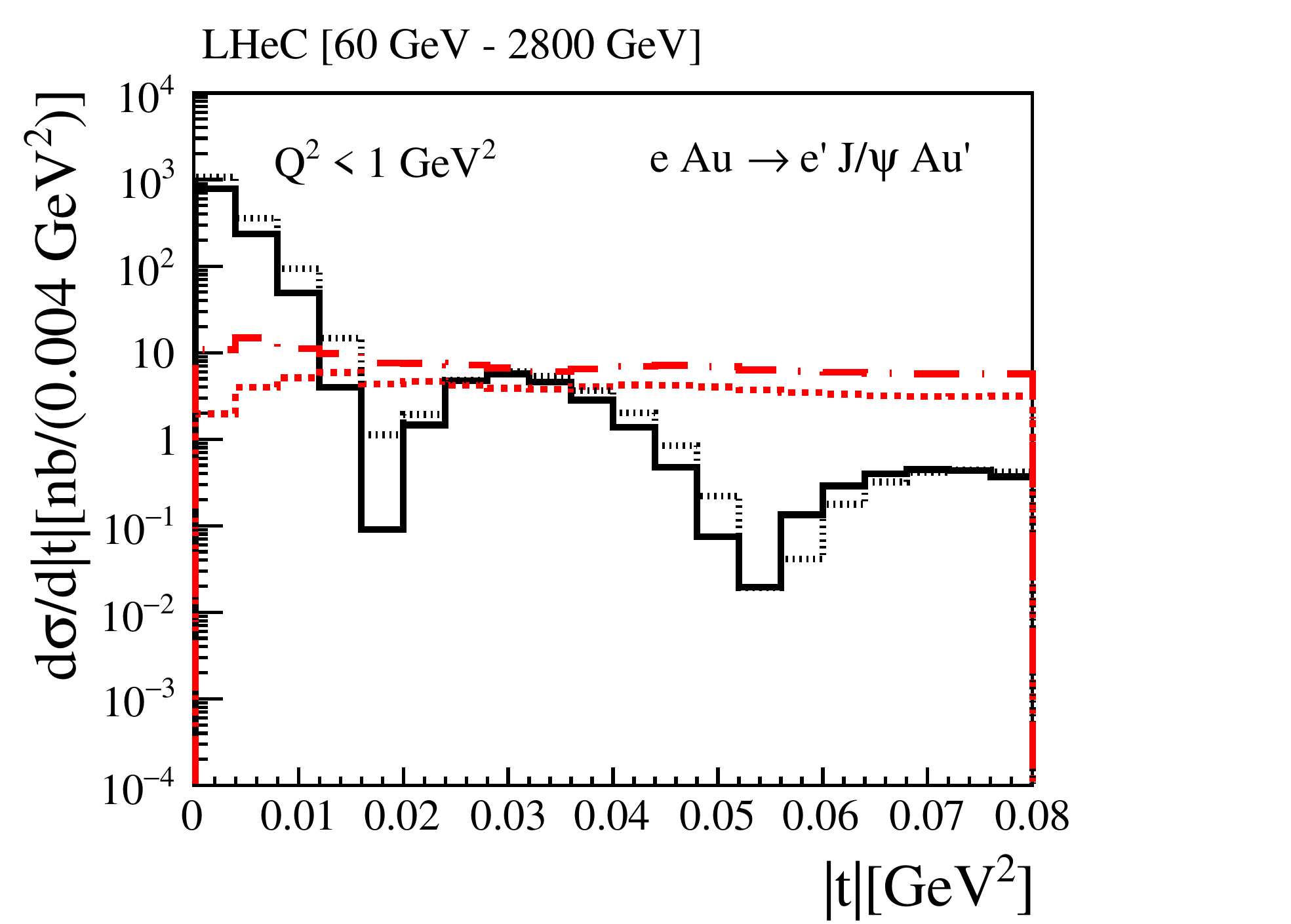}} &  {\includegraphics[width=0.35\textwidth]{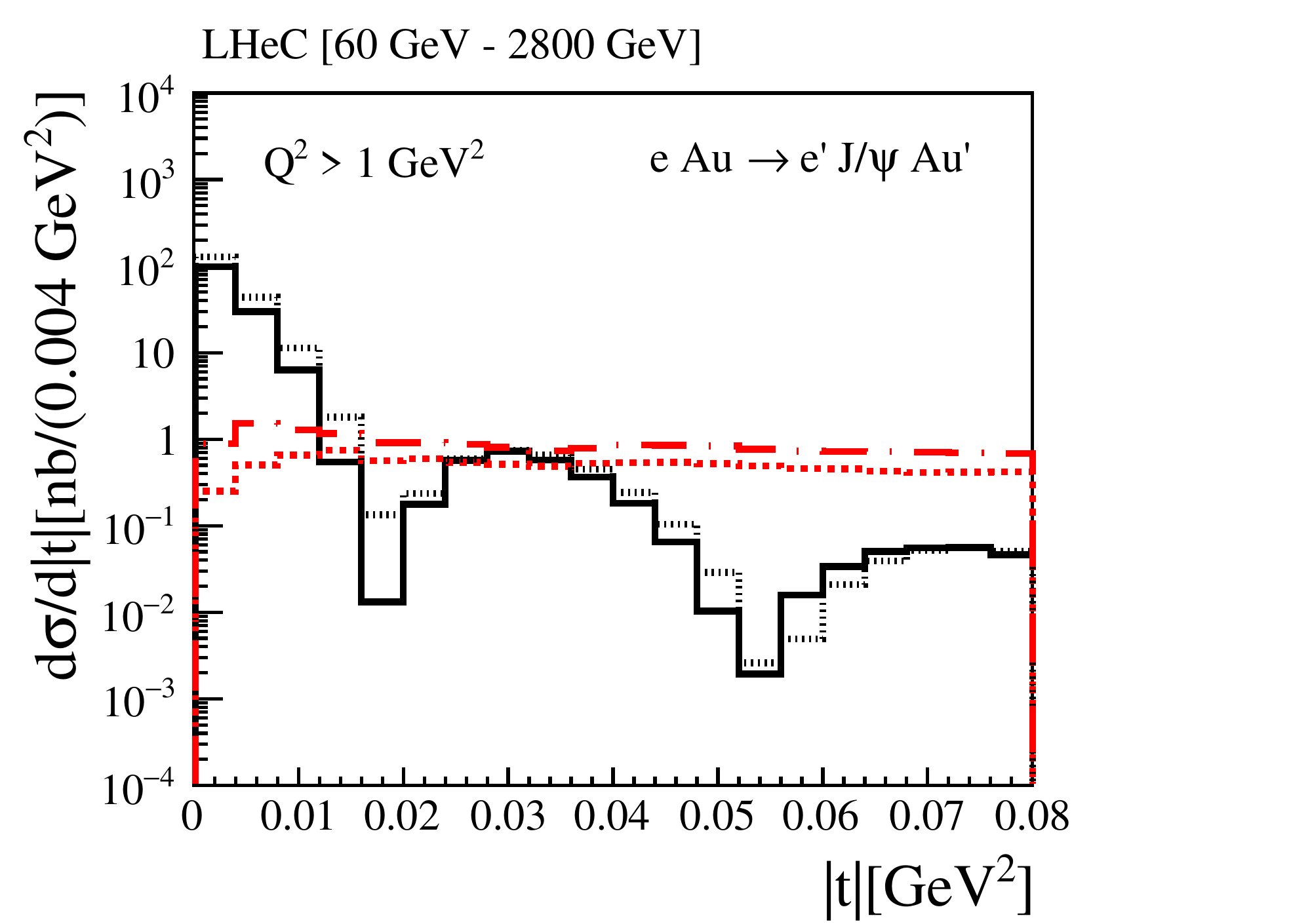}} &
{\includegraphics[width=0.35\textwidth]{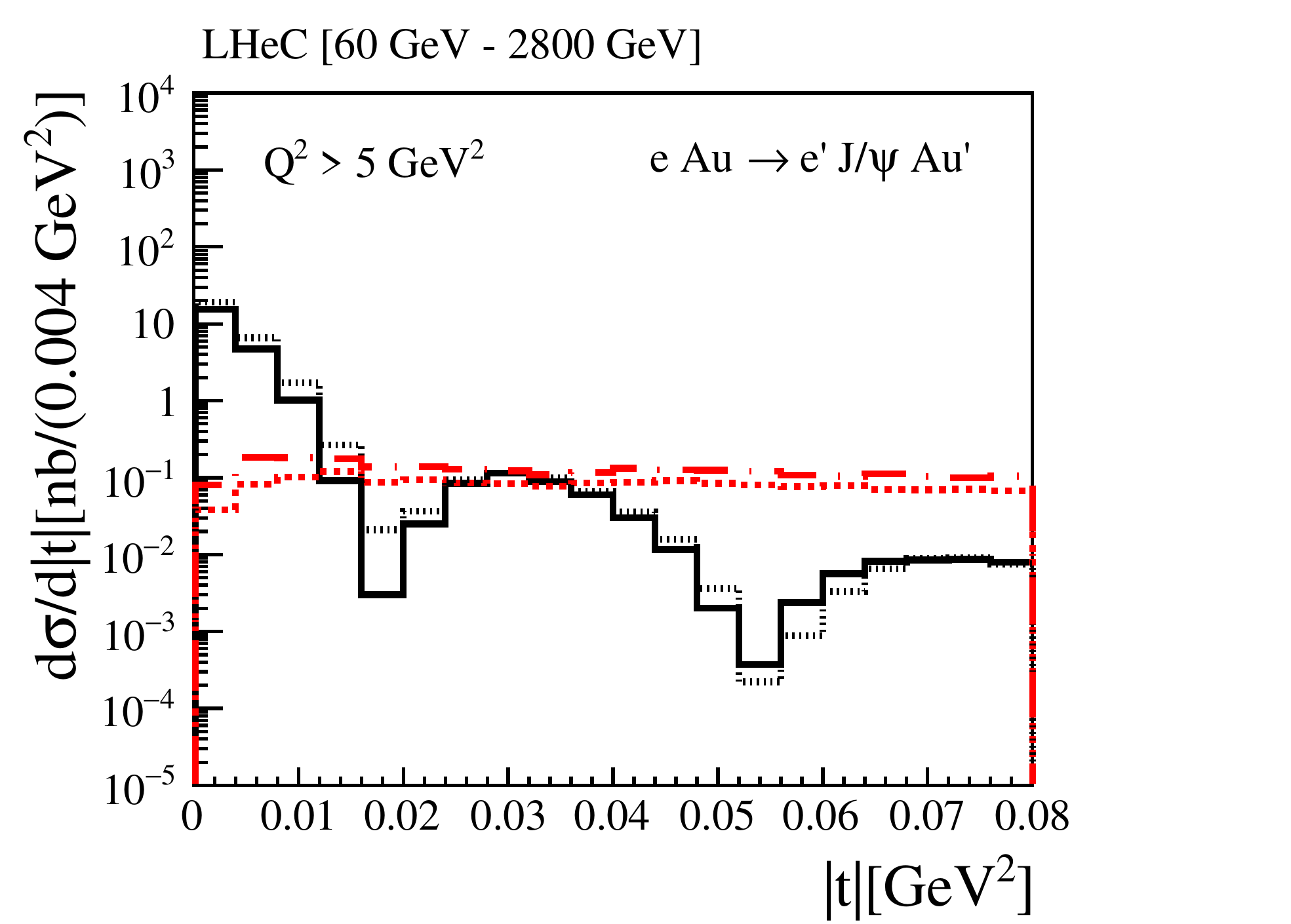}} \\
 {\includegraphics[width=0.35\textwidth]{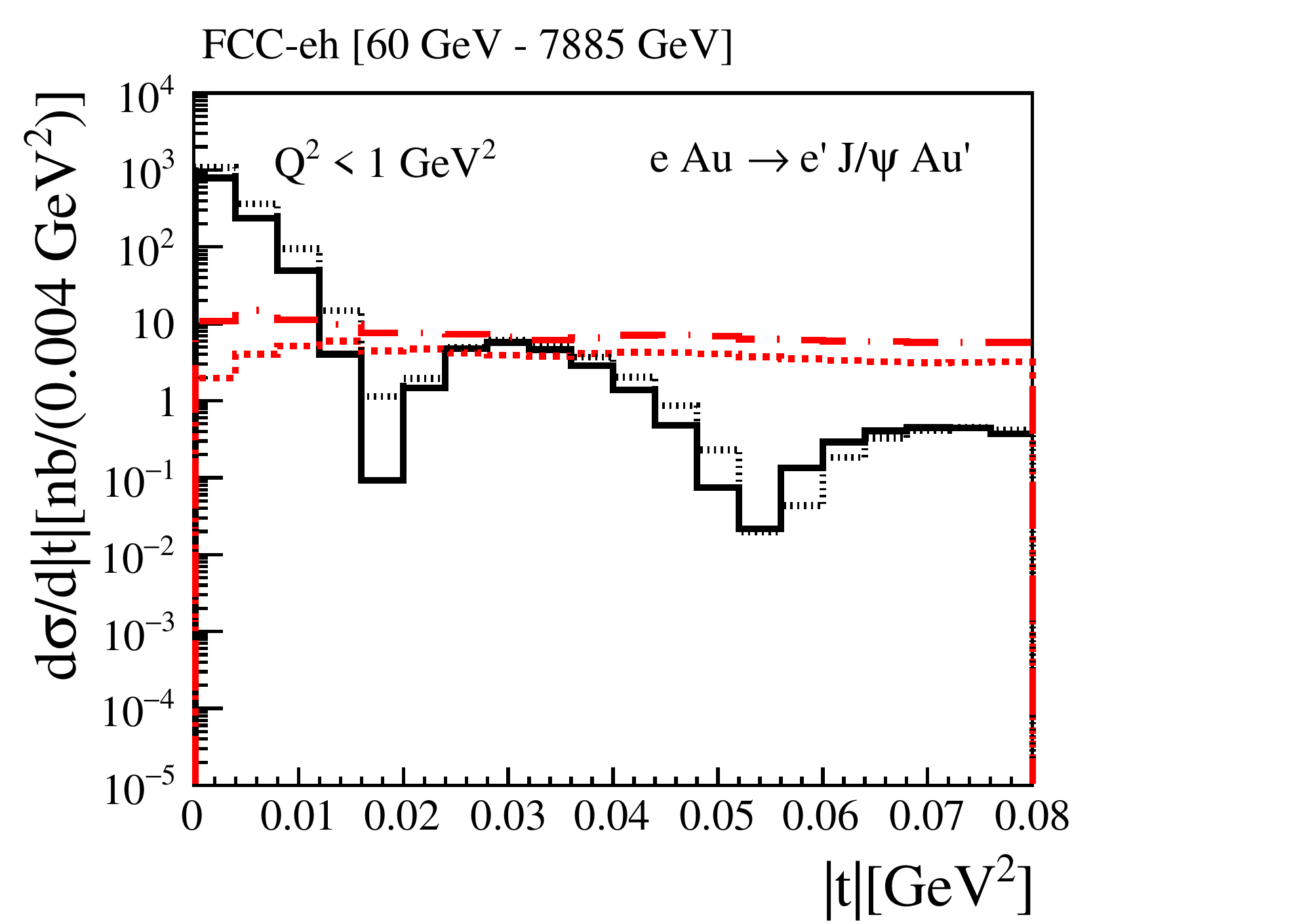}} & 
{\includegraphics[width=0.35\textwidth]{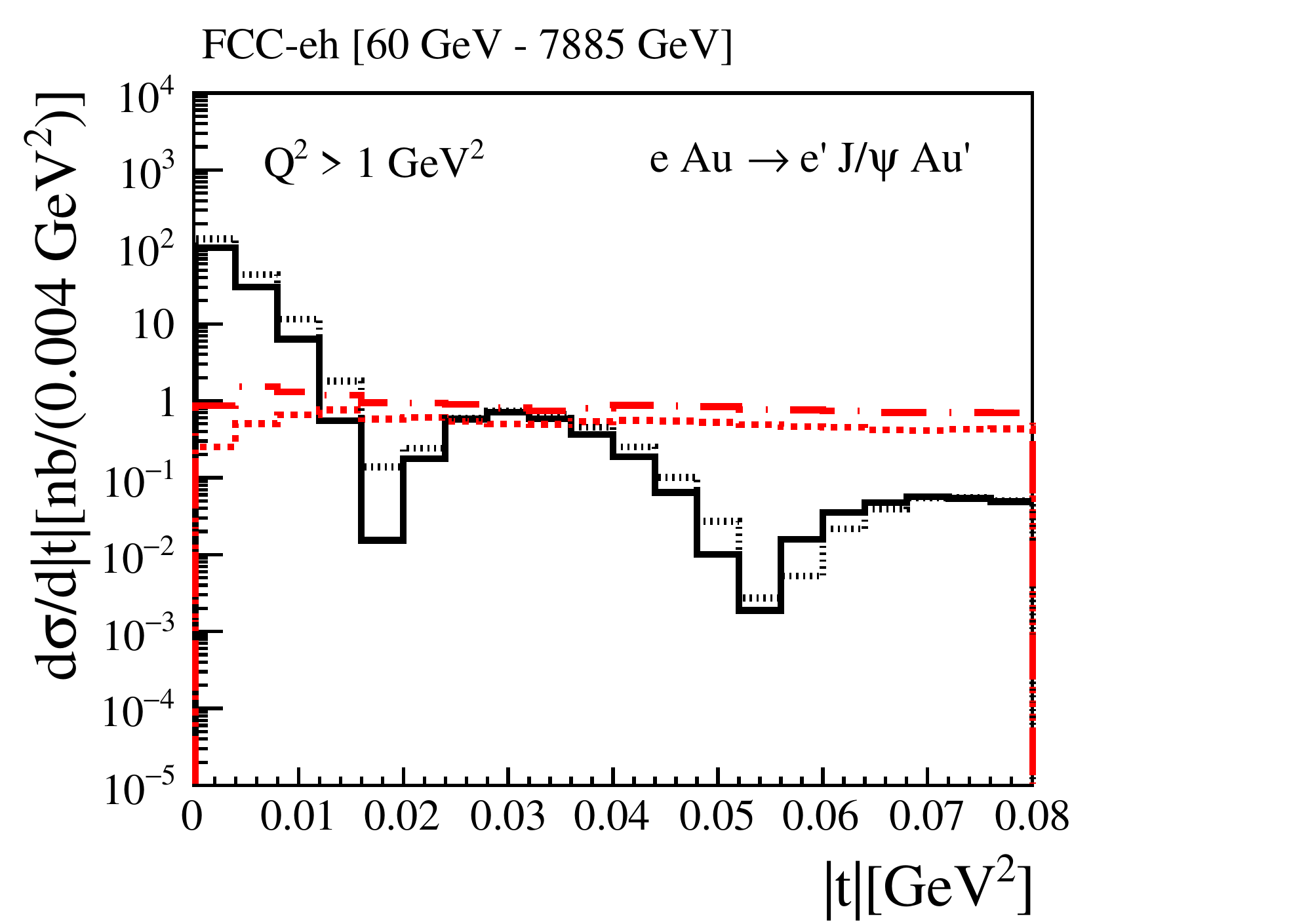}} & {\includegraphics[width=0.35\textwidth]{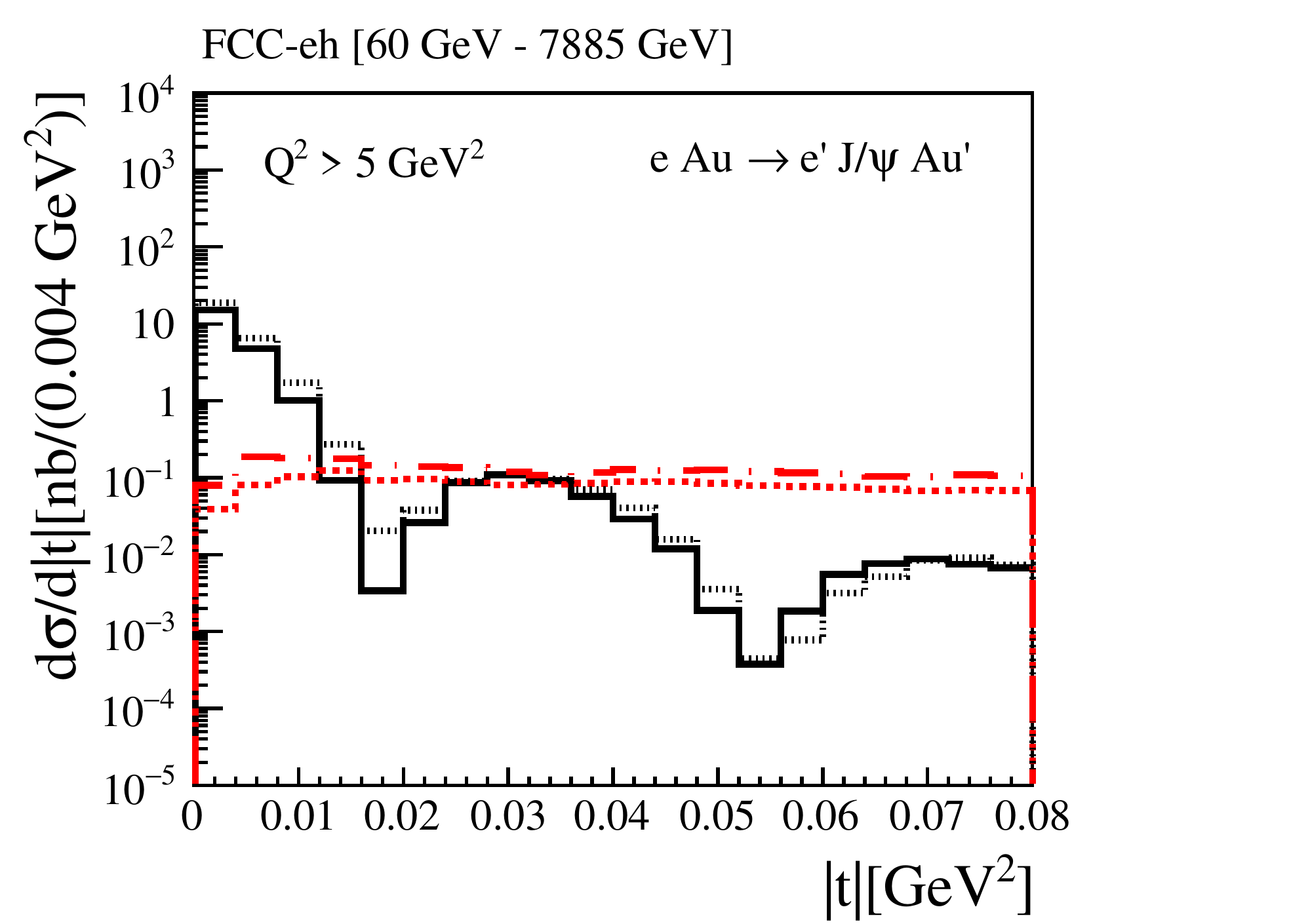}}
\end{tabular}                                                                                                                       
\caption{Transverse momentum distributions for the coherent and incoherent $J/\Psi$ photoproduction in $eAu$ collisions for the  EIC (upper panels), LHeC (middle panels) and FCC - $eh$ (lower panels) energies, and different ranges of the photon virtuality $Q^2$.}
\label{fig:trans_jpsi}
\end{figure}

 \begin{figure}[t]
\begin{tabular}{ccc}
 {\includegraphics[width=0.35\textwidth]{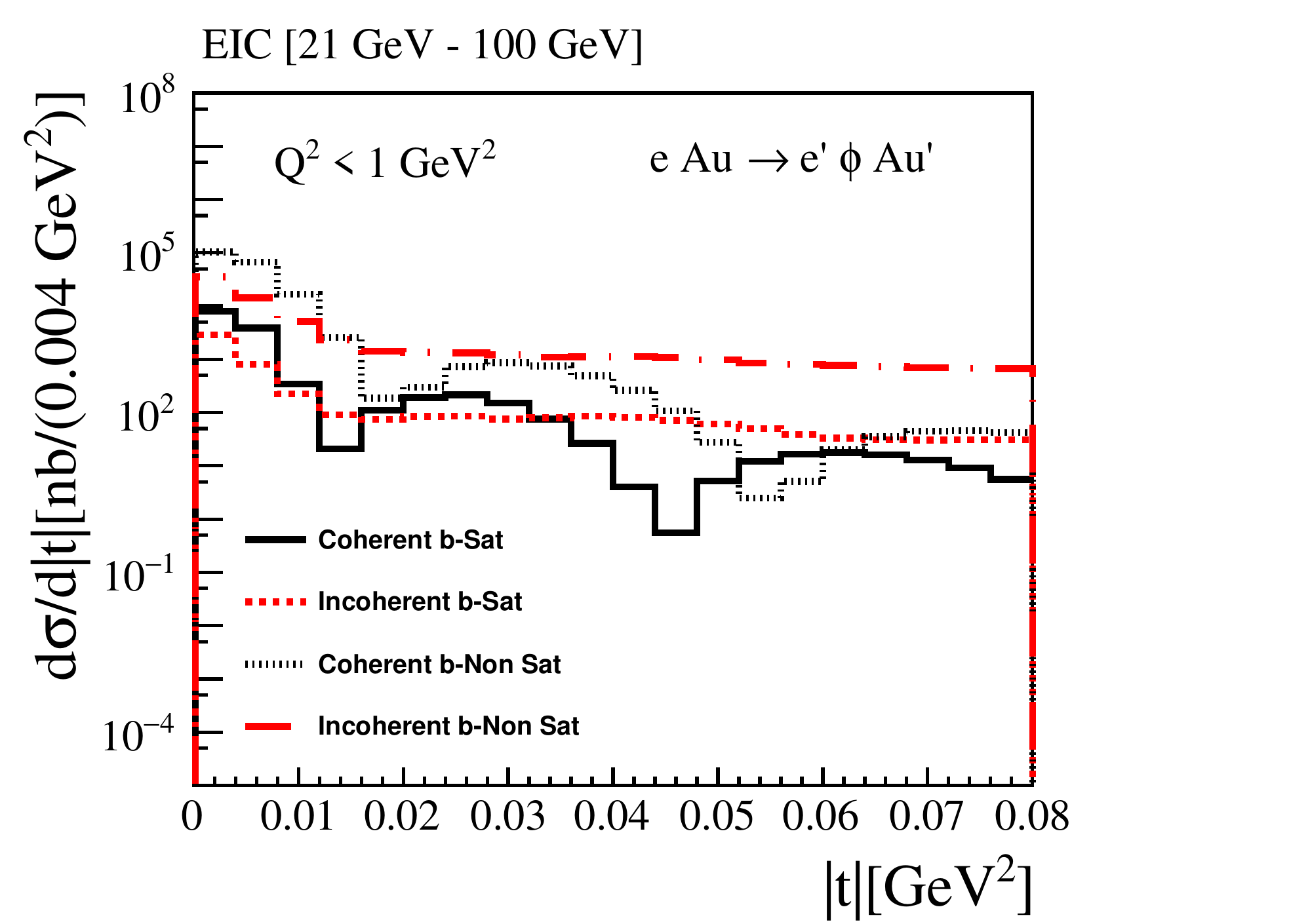}} &  {\includegraphics[width=0.35\textwidth]{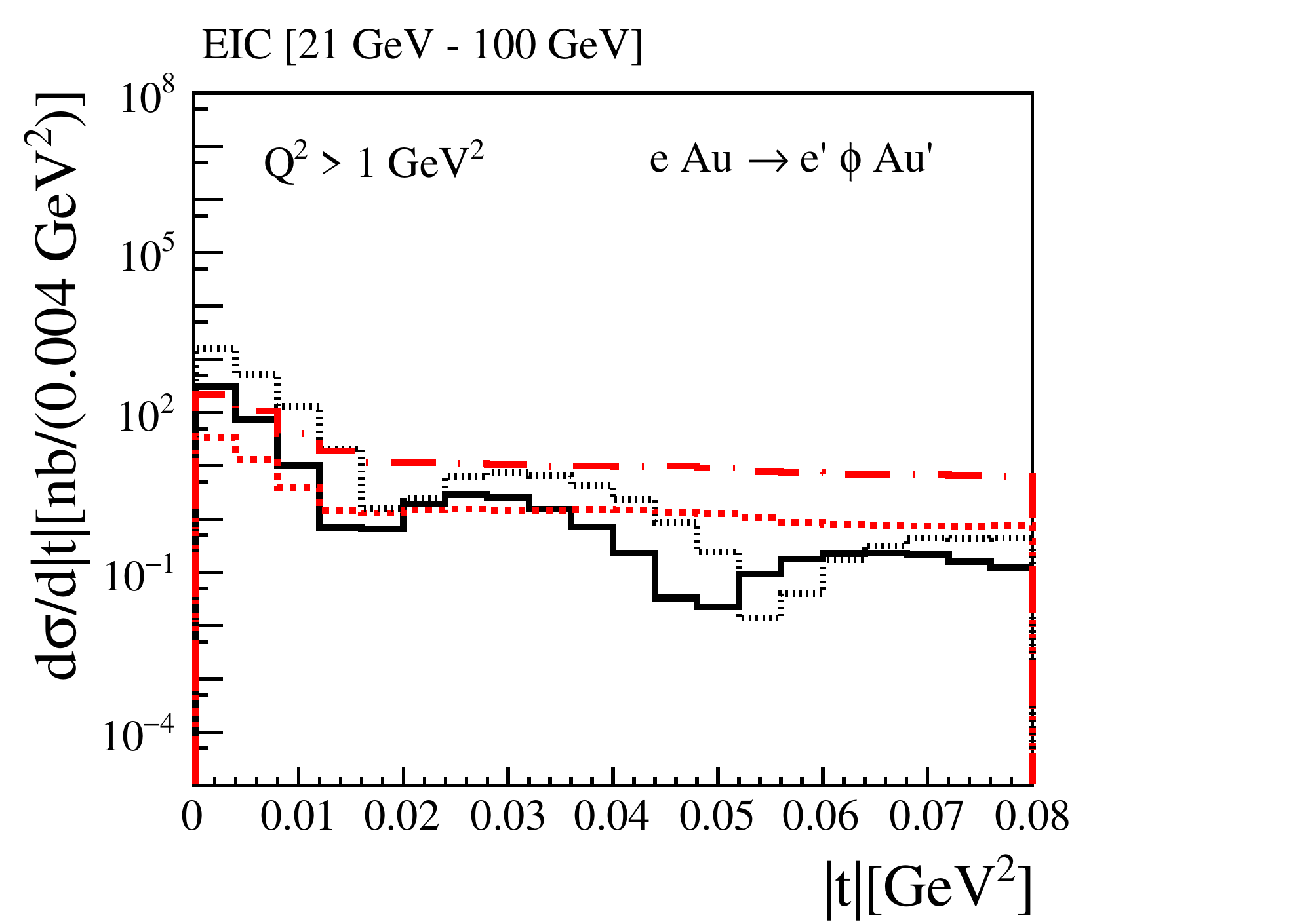}} & 
{\includegraphics[width=0.35\textwidth]{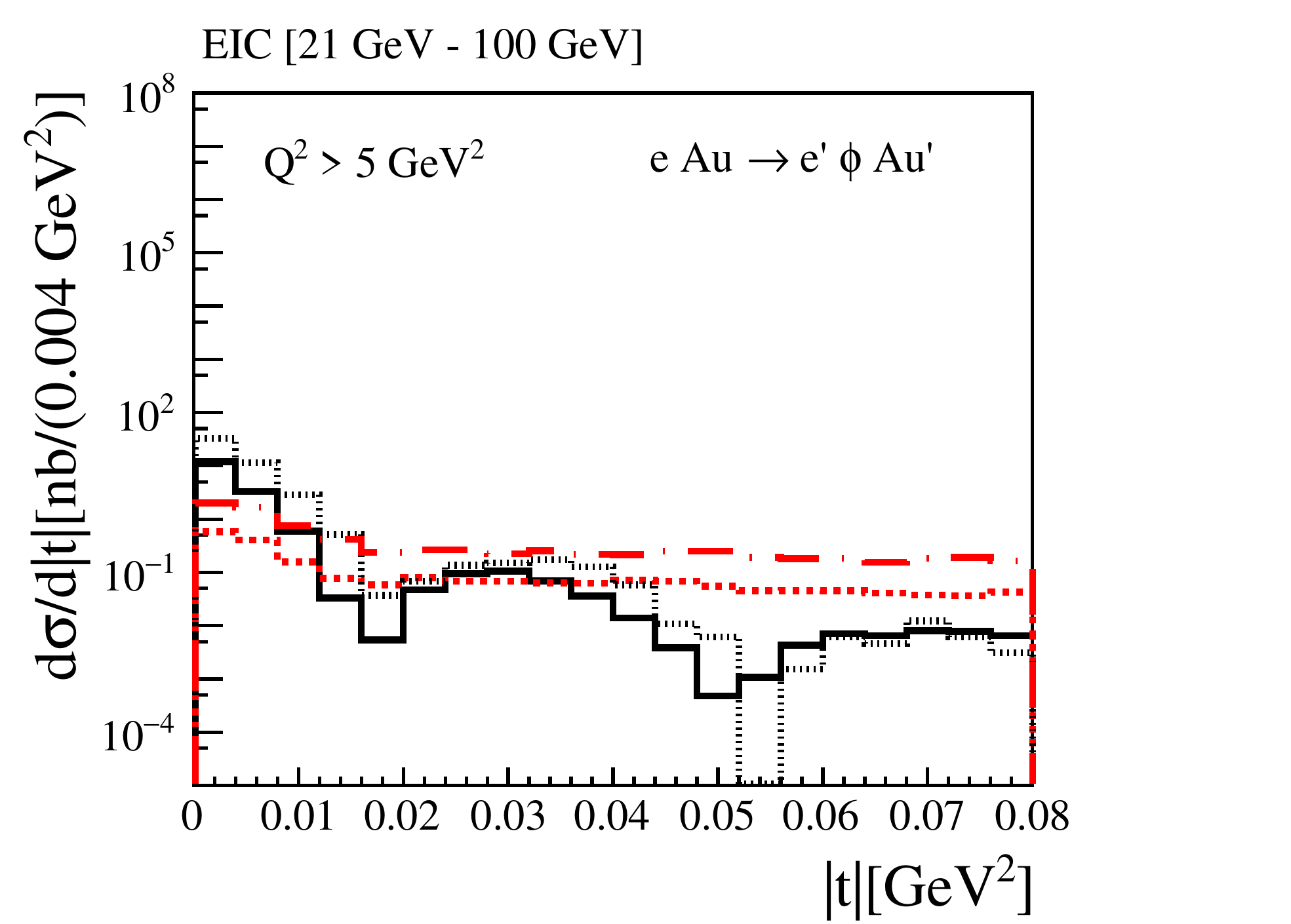}} \\
 {\includegraphics[width=0.35\textwidth]{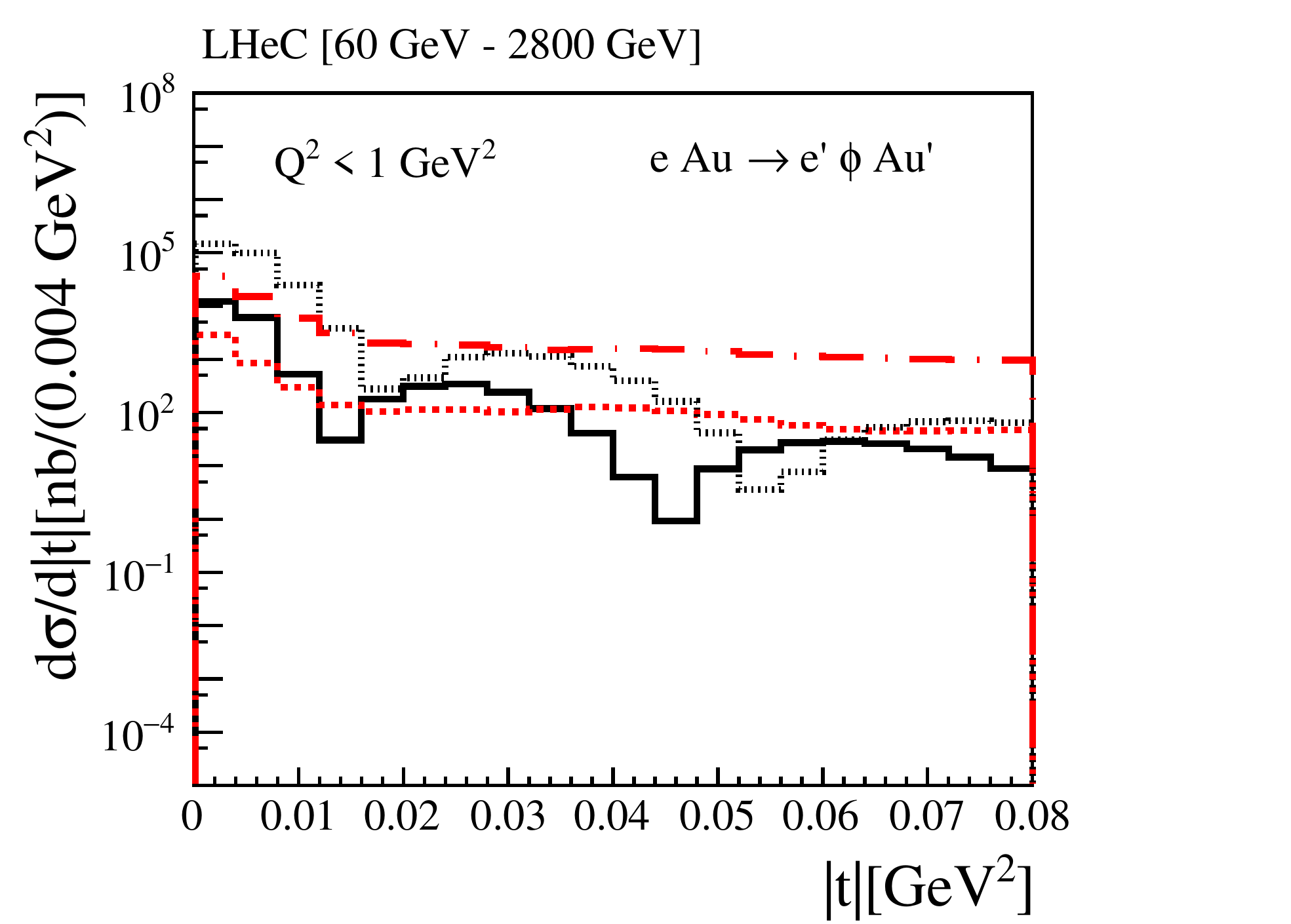}} &  {\includegraphics[width=0.35\textwidth]{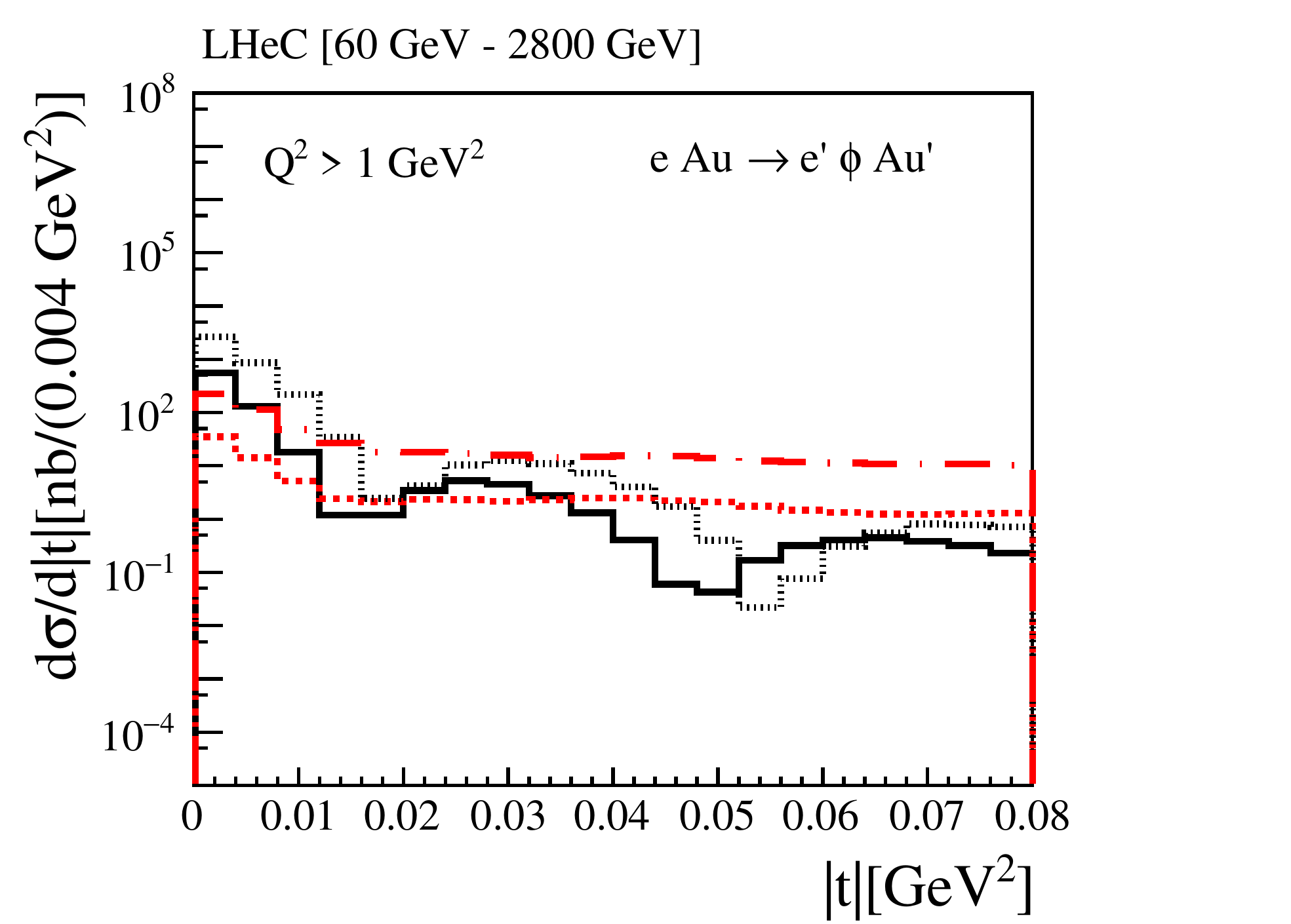}} & 
{\includegraphics[width=0.35\textwidth]{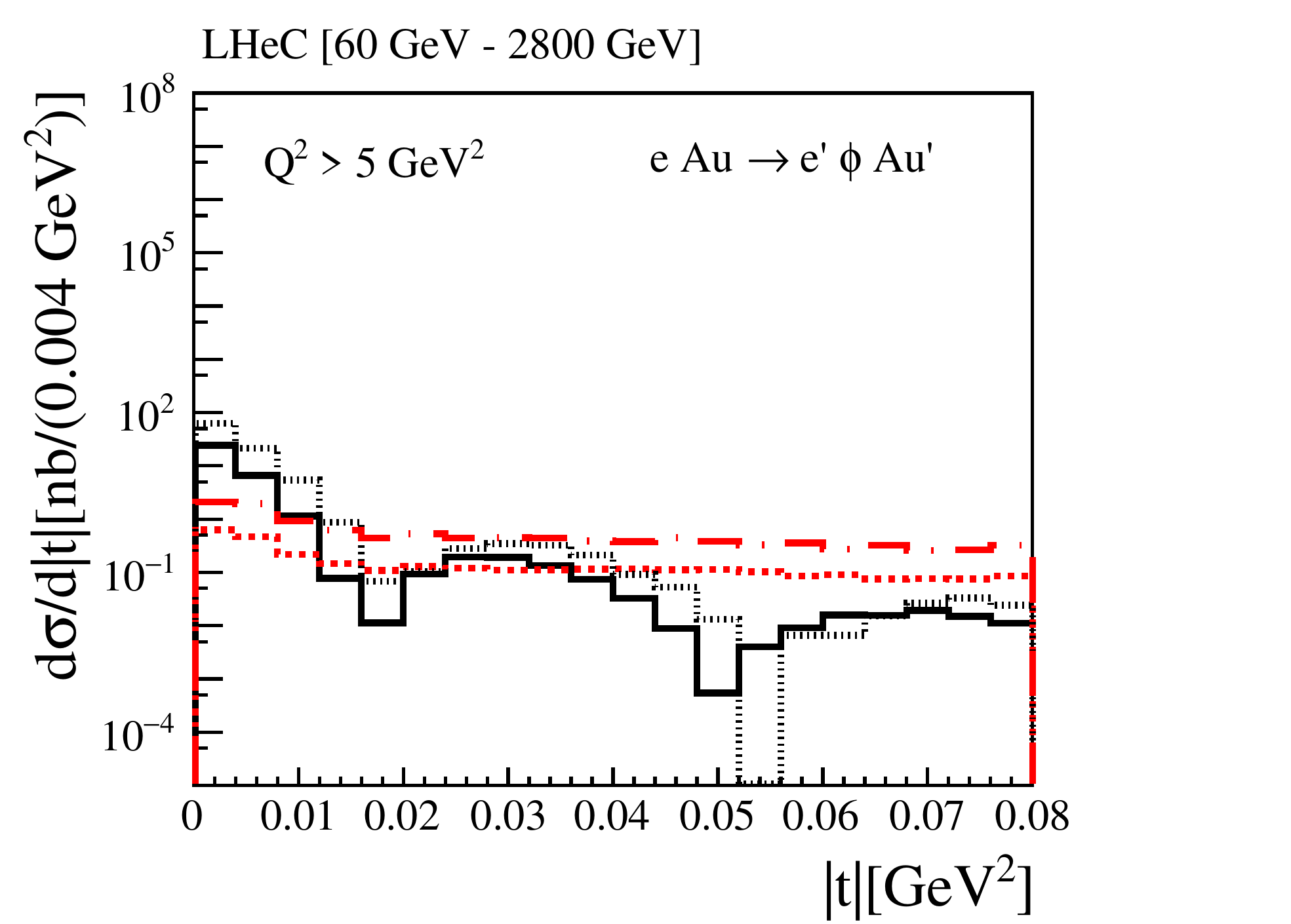}} \\
 {\includegraphics[width=0.35\textwidth]{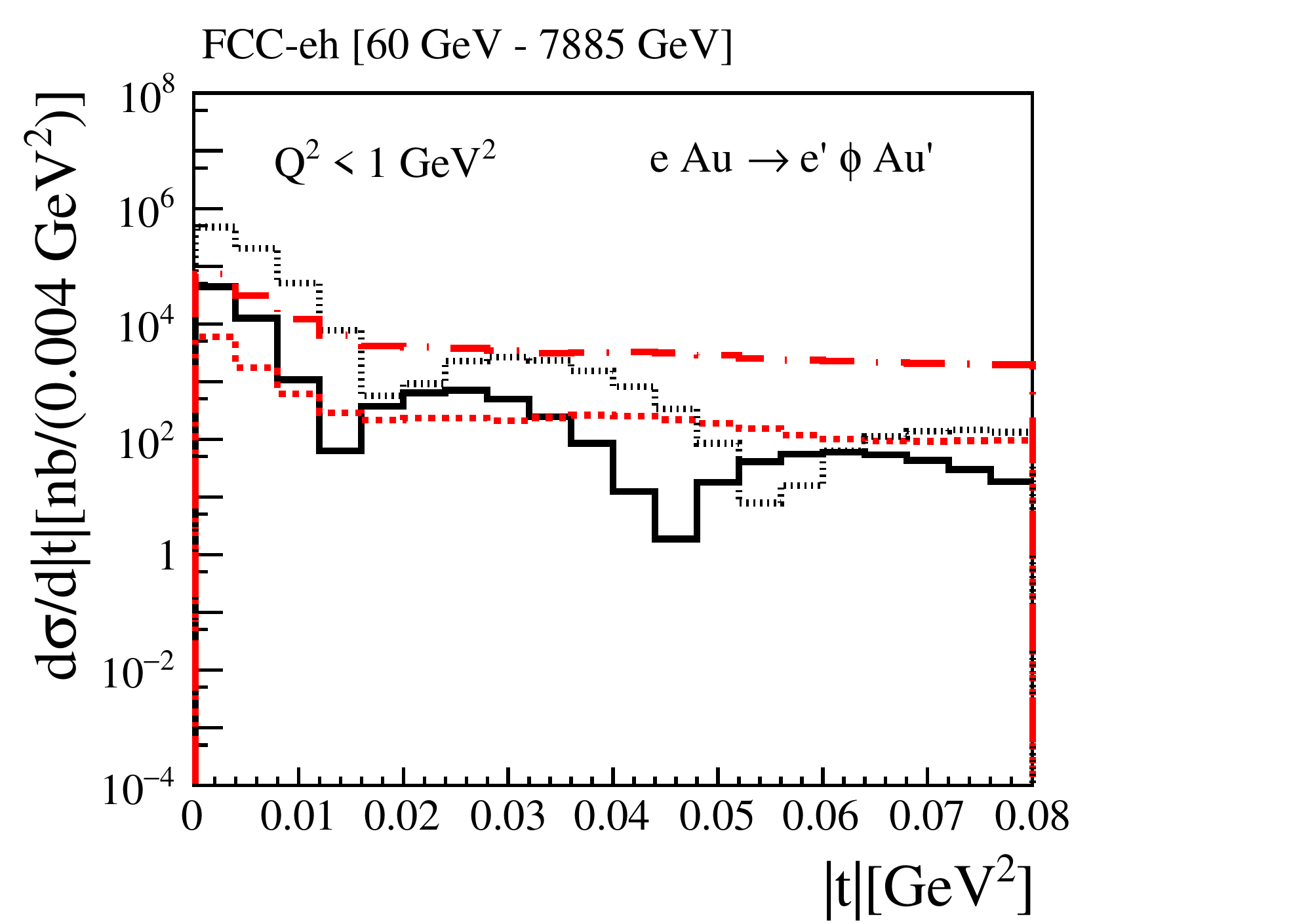}} &  {\includegraphics[width=0.35\textwidth]{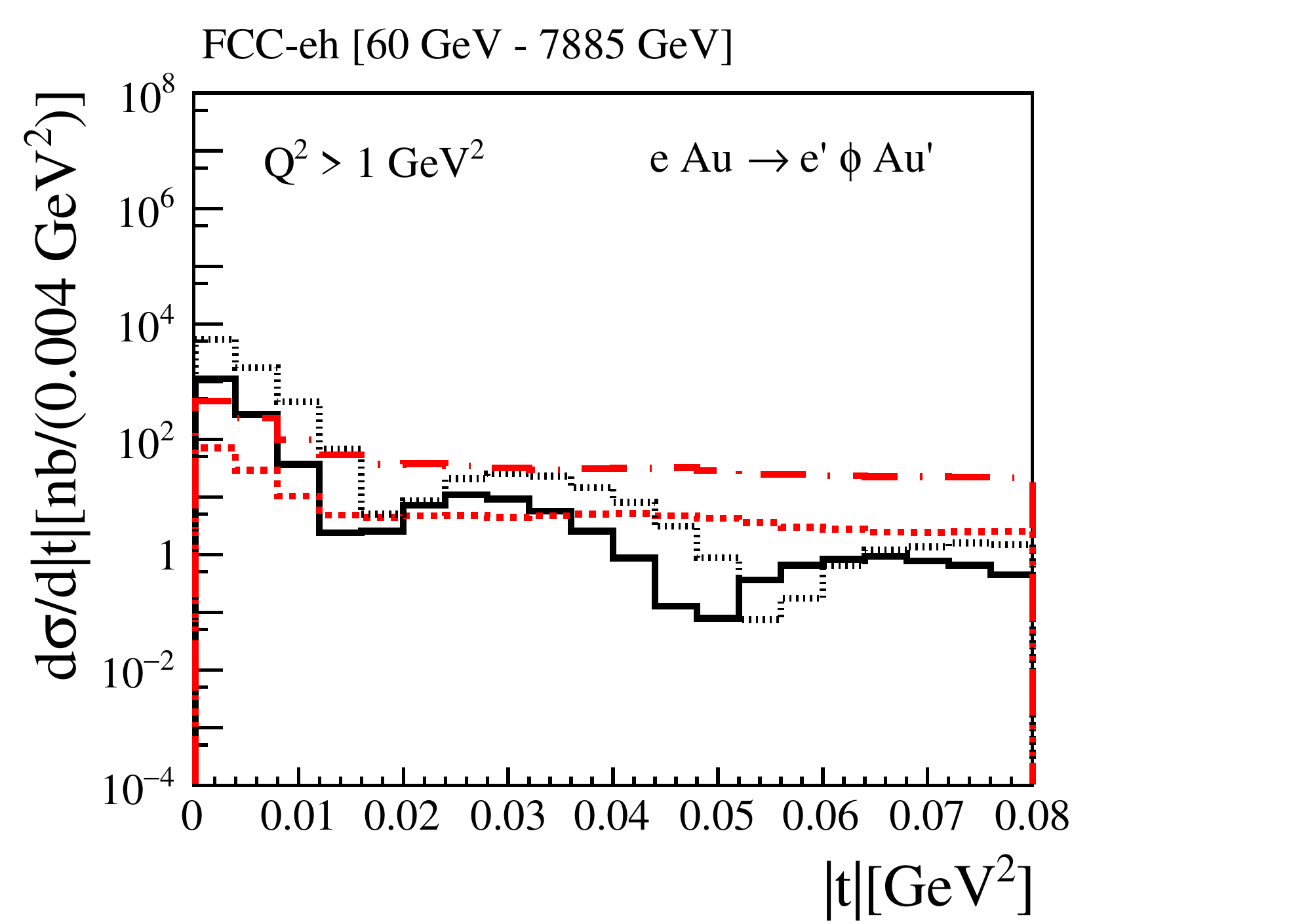}} & 
{\includegraphics[width=0.35\textwidth]{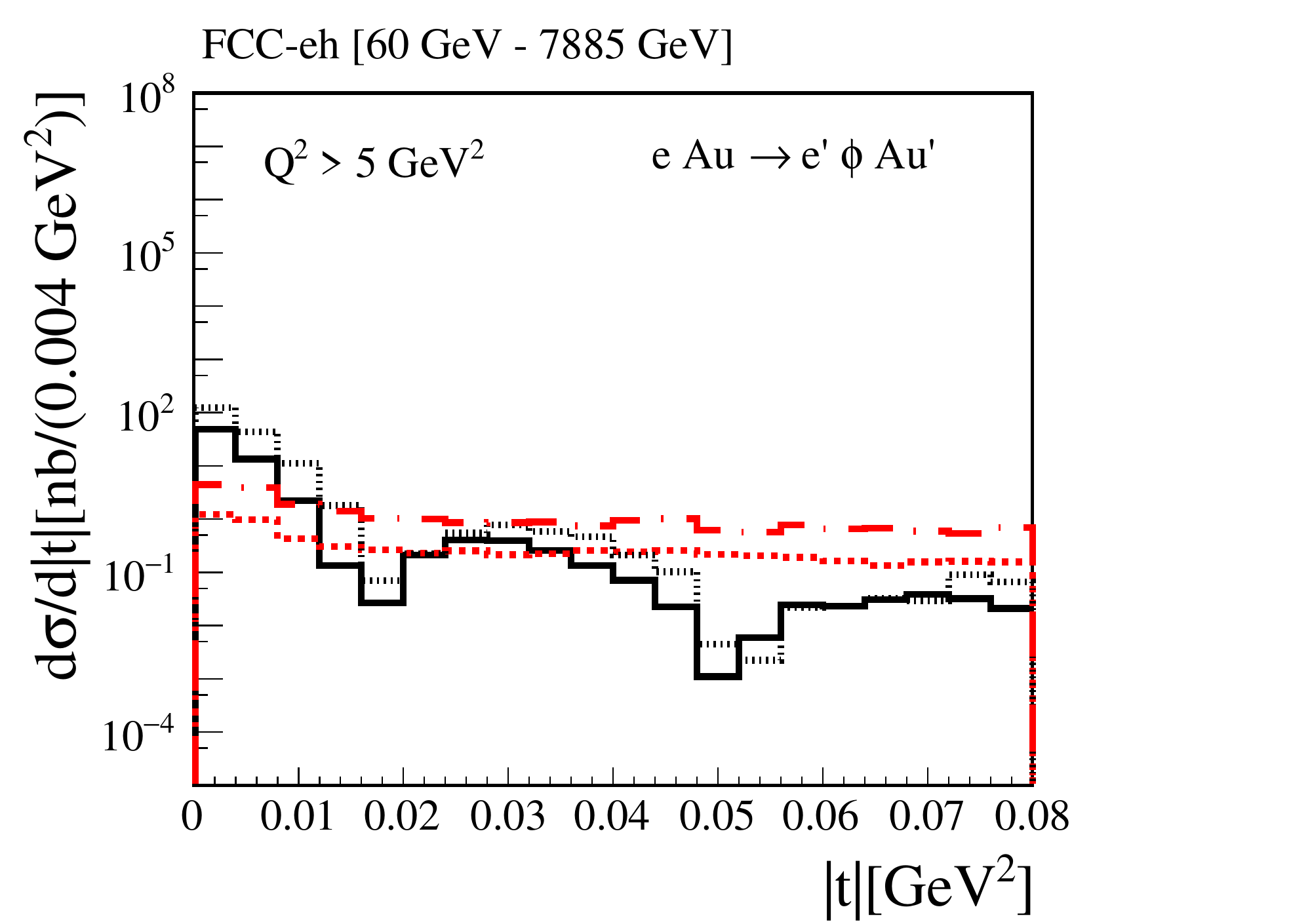}}
\end{tabular}                                                                                                                       
\caption{Transverse momentum distributions for the coherent and incoherent $\phi$ photoproduction in $eAu$ collisions for the  EIC (upper panels), LHeC (middle panels) and FCC - $eh$ (lower panels) energies, and different ranges of the photon virtuality $Q^2$.}
\label{fig:trans_phi}
\end{figure}

An important alternative to probe the QCD dynamics at high energies and provide  information about the spatial distribution of the gluons in the target and about fluctuations  of the  color fields is the study of the  squared  momentum transfer ($t$) distribution \cite{armestoamir,Diego1,Diego2}, which allow us to separate the coherent and incoherent processes.  Our predictions are presented in Figs. \ref{fig:trans_jpsi} and \ref{fig:trans_phi} for the $J/\psi$ and $\phi$ production, respectively. We present predictions for distinct ranges of the photon virtuality and different colliders.  One has that  the coherent production dominates at small - $|t|$ and the incoherent ones at large values of the momentum transfer, in agreement with previous results  \cite{Diego1,cepila}. Such behaviours are expected, since  the probability that  the nucleus breaks up becomes larger when the momentum kick given to the nucleus is increased and becomes zero for $|t| \rightarrow 0$, where excited states cannot be produced. Moreover, one has that  the  coherent cross sections clearly exhibit the typical diffractive pattern, being characterized by a sharp forward diffraction peak, while  the incoherent one is characterized by a flat $t$ - dependence.
We can see that the saturation effects reduce the normalization of the incoherent predictions, with the difference between the b-Sat and b-Non Sat predictions increasing with the energy and being larger for the $\phi$ production. Moreover, such difference also decreases at larger $Q^2$.
For the coherent case, one also has that the normalization is suppressed by the saturation effects, with the impact being larger for the $\phi$ production. As already observed in Refs. \cite{Diego1,Diego2}, the position of the dips in the coherent predictions is sensitive to the presence of the saturation effects.  Our results for the $J/\psi$ production indicate that the position of the second dip is more dependent on description of the QCD dynamics, with the predictions becoming more distinct at larger energies. However, the difference between the b-Sat and b-NonSat predictions is small for this final state, which implies that the coherent $J/\psi$ production is not ideal  to discriminate between these two scenarios. In contrast, our results for the $\phi$ production, presented in Fig. \ref{fig:trans_phi}, demonstrate that this final state is very sensitive to the saturation effects, with the position of the dips being strongly dependent on the model used to describe the dipole -- nucleus cross section. One also has that the difference between the predictions decreases for larger photon virtualities. Such result is expected, since for large $Q^2$, the hard scale $\mu$ becomes larger than the nuclear saturation $Q_{s,A}$ and the process in this kinematical range is dominated by the linear QCD dynamics. Our results for the $\phi$ production indicate that a future experimental analysis of this final state will be able to constrain the presence of the saturation effects, as well as to probe the transition between the linear and non-linear regimes of the QCD dynamics. 

Finally, let's summarize our main conclusions. Future electron - ion collisions will allow to study the high - gluon density regime of the QCD, where the contribution of non - linear (saturation) effects are expected to  determine the behavior of the inclusive and exclusive observables. In this letter we have investigate the impact of these effects on the  exclusive vector meson production. We have focused in the coherent and incoherent $\phi$ and $J/\psi$ production, which probe different regimes of the QCD dynamics. We have estimated the cross sections and transverse momentum distributions for the kinematical ranges that will be probed by the EIC, LHeC and FCC - $eh$,   considering  
the possible states of nucleon configurations in the nuclear wave function and taking into account of the non - linear corrections to the QCD dynamics. Moreover, a comparison with the results derived disregarding these corrections was also presented. 
We have  demonstrated that the event rates of these processes are very large and that the $\phi$ production is very sensitive to saturation effects. In particular, these results indicate that the experimental analysis of the transverse momentum distribution is  useful to discriminate between different approaches for the QCD dynamics as well to improve our description of the gluon saturation effects. Finally, our results indicate  that a future experimental analysis of the coherent and incoherent processes will be useful to improve our understanding of the QCD dynamics at high energies.

\begin{acknowledgments}
VPG acknowledge useful discussions about coherent and incoherent interactions with Jan Cepila, Michal Krelina and Wolfgang Schafer.
This work was  partially financed by the Brazilian funding
agencies CNPq, CAPES,  FAPERGS, FAPERJ and INCT-FNA (processes number 
464898/2014-5 and 88887.461636/2019-00).
\end{acknowledgments}

\hspace{1.0cm}

\end{document}